\documentclass[a4paper,12pt]{article}
\usepackage[utf8]{inputenc}
\usepackage[T1]{fontenc}
\usepackage{subfig}
\usepackage{hyperref}
\usepackage{color,soul}
\usepackage[normalem]{ulem}
\usepackage{amsmath,amssymb,amsfonts}
\usepackage{cite}
\usepackage[compress,numbers]{natbib}
\usepackage[pdftex,dvipsnames]{xcolor}
\usepackage{graphicx}
\usepackage{caption}
\usepackage{titlesec}
\usepackage{mathpazo}
\usepackage{txfonts}

\numberwithin{equation}{section}
\titleformat{\section}{\normalfont\sffamily\Large\bfseries}{\thesection}{1em}{}
\titleformat{\subsection}{\normalfont\sffamily\large\bfseries}{\thesubsection}{1em}{}

\newcommand{\Walberla}{\textsc{waLBerla}}
\DeclareMathAlphabet{\mathpzc}{OT1}{pzc}{m}{it}
\newcommand{\pe}{$\mathpzc{pe}$}

\begin{document}
\begin{center}
{\Large{{\bf{\textsf{Microswimming with inertia}}}}}
\end{center}
\begin{center}
\vskip6pt
Jayant Pande$^{*1,2}$, Kristina Pickl$^{*2,3}$, Oleg Trosman$^{1,2}$, Ulrich R{\" u}de$^{1}$ and Ana-Sun\v{c}ana\ Smith$^{2,3,4\dagger}$

\vskip12pt

$^2${\footnotesize{\emph{Cluster of Excellence: EAM, Friedrich-Alexander University Erlangen-N\"urnberg, N\"agelsbachstra\ss e 49b, 91054 Erlangen, Germany}}}\\
$^3${\footnotesize{\emph{PULS Group, Department of Physics, Friedrich-Alexander University Erlangen-N\"urnberg, N\"agelsbachstra\ss e 49b, 91054 Erlangen, Germany}}}\\
$^1${\footnotesize{\emph{Chair for System Simulation, Friedrich-Alexander University Erlangen-N\"urnberg, Cauerstra\ss e 11, 91058 Erlangen, Germany}}}\\
$^4${\footnotesize{\emph{Division of Physical Chemistry, Ruđer Bošković Institute, Bijeni\v{c}ka cesta 54, Zagreb, Croatia}}}\\
\vskip4pt
$^*${\footnotesize{These authors contributed equally to the work.}}\\
$^\dagger${\footnotesize{Author for correspondence. E--mail:\ \href{mailto:smith@physik.fau.de}{smith@physik.fau.de}. }}
\vskip18pt
\end{center}

\begin{abstract}
Microswimmers, especially in theoretical treatments, are generally taken to be completely inertia-free, since inertial effects on their motion are typically small and assuming their absence simplifies the problem considerably. Yet in nature there is no discrete break between swimmers for which inertia is negligibly small and for which it is detectable. Here we study a microswimming model for which the effect of inertia is calculated explicitly in the regime of transition between the Stokesian and the non-Stokesian flow limits, which we term the intermediate regime. The model in the inertialess limit is the bead-spring swimmer. We first show that in the intermediate regime a mechanical microswimmer exhibits damped inertial coasting like an underdamped harmonic oscillator. We then calculate analytically the swimmer's velocity by including a mass-acceleration term in the equations of motion which are otherwise based on the Stokes flow. We show that this hybrid treatment combining aspects of underdamped and overdamped dynamics provides an accurate description of the motion in the intermediate regime, as verified here by comparison to simulations using the lattice Boltzmann method, and is a significant improvement over the results from the inertialess theory when either the mass of the swimmer or the forces driving its motion is/are large enough.
\end{abstract}

\section{Introduction}
\label{sec:Introduction}

The motion of microswimmers, whether natural ones like \emph{E.\ coli}~\cite{Berg:2004:Springer} and \emph{C.\ reinhardtii}~\cite{Guasto:2010:PRL} or artificial ones like various proposed micro-machines~\cite{Purcell:1977:AmJPhys, Najafi:2004:PRE, Dreyfus:2005:Nature, Avron:2005:NewJPhys, Baraban:2012:SoftM}, is dominated by the viscous forces exerted by the surrounding fluid, with the inertia of the swimmers playing practically no role. For these organisms the Reynolds number ($Re$) is typically very small, usually $10^{-2}$ or smaller. Because of this, a common approach in the literature to model their motion is to assume that the $Re$ vanishes~\cite{Yeomans:2014:EPJST, Happel:1965:P-H, Koch:2011:AnnuRevFluidMech, Lauga:2009:RPP, Elgeti:2015:RepProgPhys}. In this limit the non-linear terms in the Navier-Stokes equation for the fluid flow can be eliminated, resulting in the Stokes equation, which makes the analysis of the motion much easier~\cite{Happel:1965:P-H}. The swimmers are then said to be in the ``Stokes regime''.

Larger organisms such as small insects swim in what may be termed the ``intermediate regime'', at which both the viscous forces and the inertial forces are important in directing the motion~\cite{Ngo:2014:JExpBiol}. Their Reynolds numbers typically range between 1 and a few hundred. Finally, organisms on the scale of meters, such as large fish, swim at $Re \gtrsim 10^3$, where the inertial forces dominate the motion, and the coasting of the swimmer during the non-active part of each cycle is an integral part of the swimming strategy~\cite{Childress:1981:CUP}. This is the domain of the ``non-Stokes regime''.

The above demarcation of the three regimes is only loosely defined, with the specific details of a swimmer controlling its Reynolds number as well as the importance of inertial effects in determining its motion. In practice $Re < 1$ is often taken to be a sufficient condition for assuming the Stokes regime, but the validity of such an assumption is not established. Moreover, inherently non-Stokesian methods such as lattice Boltzmann (LB) simulations are often used to simulate microswimmers with $Re < 1$, and any inertial effects if present in the simulation are typically ignored in the analysis.

In this work we consider the motion of microswimmers beyond the bounds of the Stokes regime, \emph{i.e.}\ when effects of inertia are no longer negligible. As discussed above, the assumption of Stokes flow simplifies the swimming problem a lot, as the inconvenient non-linear and time-dependent terms in the Navier-Stokes equation disappear. This assumption results unavoidably in some non-physical effects, such as a direct proportionality of the swimming speed to the applied force and an instantaneous response of the fluid, up till infinity, to any disturbance within it. These effects can usually be ignored, firstly in the interest of the linearity of the resulting Stokes equation which is quite evidently crucial for analytic treatments of the swimming problem, and secondly because these effects turn out not to cause significant errors in the flow description.

Inclusion of the inertial contributions to the flow is a very daunting task in general. To bring it within the realm of tractability, we restrict our attention to the lowest order effects of inertia on our swimmer. In other words, we study the swimming regime where inertial or non-Stokesian effects first emerge and impart quantitative, and possibly qualitative, differences to the motion when compared to that in purely Stokes flow. Our approach will be to take inertia into account modifying the methods of a Stokes flow calculation. We will do this in two ways. Firstly, we will provide a heuristic scheme for identifying the non-Stokesian regime of motion by considering the relaxation of a swimmer from an initially stably-moving configuration. In the Stokesian approximation, this relaxation is instantaneous, and the degree of non-instantaneity in it will speak to the non-Stokesian aspect of the motion. We will show that the larger the mass of the swimmer, the slower does it have to be for the relaxation to display an inertial component. 
Secondly, we will perform a full calculation of the swimming velocity of our bead spring swimmer model with non-negligible inertia, by including a mass acceleration term in the equations of motion governing the swimming. This shall confirm the finding that for larger swimmer masses the inertial or the non-Stokes regime begins at smaller swimming speeds (\emph{i.e.}\ smaller driving forces). In both cases the theoretical predictions will be supported by lattice Boltzmann simulations.

\section{Swimmer model}
\label{sec:Model}
\begin{figure}[htb]
\centering
  \includegraphics[width=0.9\columnwidth]{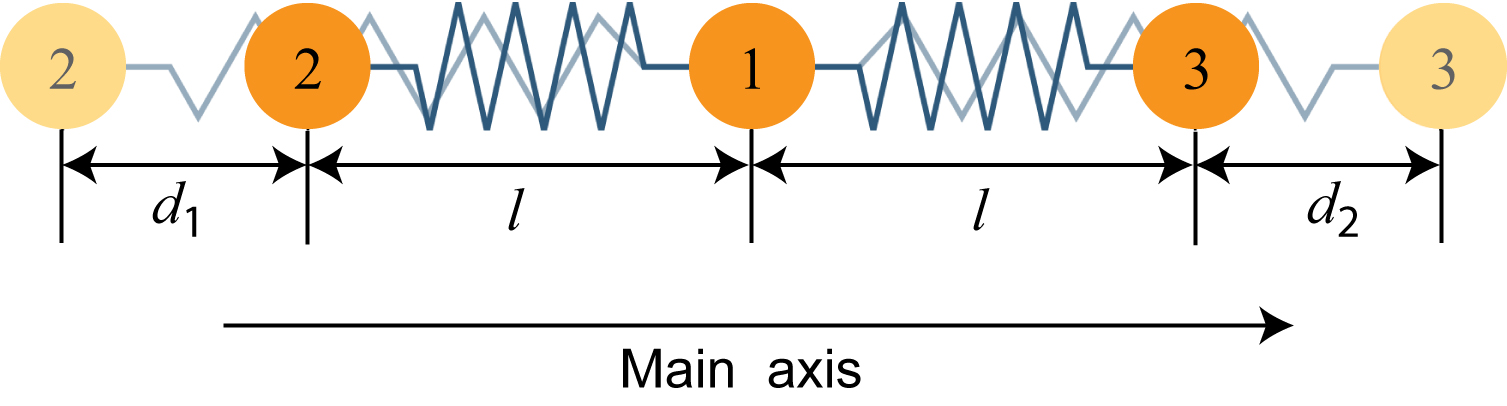}
\caption{Three-sphere swimmer model, based on \cite{Najafi:2004:PRE}.}
\label{fig:swimmer}     
\end{figure}
Our swimmer, based on the Najafi-Golestanian design~\cite{Najafi:2004:PRE}, consists of three rigid spheres of equal masses $m$ and 
radii $r$ connected by two linear springs of equal stiffness constants $k$ and mean rest lengths $l$. In the simulations, additionally, we include angular springs of high stiffness constants to ensure linear and rotational alignment of the swimmer~\cite{Pickl:2014:IOS}. Time-irreversibility in the stroke, as required by the Scallop Theorem for net propulsion at zero Reynolds number~\cite{Purcell:1977:AmJPhys}, is achieved via the sinusoidal driving forces
%
\begin{align}
\label{eq:driving1}
 \mathbf {F}_{\mathrm{d}_1}(t) &= - \left({F}_{\mathrm{d}_2}(t) + {F}_{\mathrm{d}_3}(t)\right) \hat{\mathbf z}, \\
\label{eq:driving2}
 \mathbf {F}_{\mathrm{d}_2}(t) &= - a \sin\left(\omega t \right) \hat{\mathbf z},\text{ and} \\
 \label{eq:driving3}
 \mathbf {F}_{\mathrm{d}_3}(t) &= b \sin\left(\omega t + \alpha \right) \hat{\mathbf z},\text{ for } t \ge \left(\pi - \alpha\right)/\omega,
\end{align}
%
that are applied to the centers of mass of the spheres along the main axis of the swimmer, the $x$-axis. The parameters $a$ and $b$ denote the amplitudes of the sinusoidal forces, $\omega$ is the force frequency, $T$ is the cycle period (so that $T = 2\pi/\omega$), $\alpha$ is the relative phase shift in the forces and $t$ is the instantaneous time. The force ${F}_{\mathrm{d}_3}$ on
the right sphere is kept zero for the first $\left(\pi - \alpha\right)/\omega$ time steps, so that it starts continuously from a value of zero as that aids in the stability of simulations; this has no bearing on the model itself. Due to Eq.~(\ref{eq:driving1}) the criterion of force neutrality is always satisfied. As a result of the forces, the trailing and leading arms perform sinusoidal motion with oscillation amplitudes $d_1$ and $d_2$, respectively (Fig.~\ref{fig:swimmer}).
%
\section{Simulation framework}
\label{sec:SimulationMethod}
The simulation system consists of two parts, the parallel software framework 
\Walberla{}~\cite{Feichtinger:2011:JOCS, Feichtinger:2011:PC, Koestler:2013:IT} for simulating the fluid, and the rigid body engine 
\pe{}~\cite{Iglberger:2011:MSD} for the bodies within it.

\Walberla{} is based on the lattice Boltzmann method (LBM)~\cite{Aidun:2010:AnnuRevFluidMech, Succi:2001:Clarendon} and uses a D3Q19 
model~\cite{Qian:1992:EPL} for three-dimensional space discretisation employing 19 particle distribution functions~(PDFs) 
per lattice site. For the collision operator, Ginzburg's two-relaxation time model \cite{Ginzburg:2008:CCP} is 
applied, which splits the PDFs, the used incompressible equilibrium distribution function, and the relaxation parameter 
into symmetric and antisymmetric parts.
For the simulations, the best accuracy at the solid walls is achieved with the symmetric relaxation parameter set to 
$1/\tau$ and the antisymmetric one set to $8(2-1/\tau)/(8-1/\tau)$~\cite{Ginzburg:2008:CCP}, 
where $\tau$ denotes the relaxation time.
The software utilizes the message passing interface (MPI) and is 
optimized for scalable and efficient execution on the fastest supercomputers 
available~\cite{Godenschwager:2013:ACM}.

The massively parallel rigid body engine \pe{}~\cite{Iglberger:2011:MSD} handles the dynamics of the bodies using Newton's 
equations of motion, and includes mechanisms for resolving frictional rigid body collisions and 
modelling external forces like gravity or the sinusoidal forces driving the swimmer. 
Potential contacts are handled in this work using the parallel Discrete Element Method (DEM)~\cite{Cundall:1979:DEM}. 
In addition to these soft-contact models it also supports, \textit{e.g.}, a contact resolution based on hard-contact 
models~\cite{Preclik:2015:CPM}.
The algorithm is augmented with an MPI communication strategy that can handle general pairwise
spring-damper systems~\cite{Pickl:2014:IOS}.
To avoid problems resulting from non-local communication among processes resulting from extended springs, communication
is restricted to those pairs of processes on which objects interact.
Global communication is avoided since it might lead to a deterioration of parallel performance and scalability on modern 
supercomputers.

The interactions between the swimmer and the fluid, as well as between different bodies within the swimmer, are 
modelled by a four-way coupling scheme~\cite{Pickl:2012:JOCS, Pickl:2014:IOS, Goetz:2010:SC10}. The rigid bodies of the swimmer 
overlap with the cells of the LBM grid and are marked as obstacles within the fluid. Those that interface to the fluid 
are specified to have a moving boundary condition~\cite{Ladd:1994:JFMa, Ladd:1994:JFMb, Yu:2003:PAS}. The fluid couples to a rigid 
body via the momentum exchange method~\cite{Ladd:1994:JFMa, Ladd:1994:JFMb, Iglberger:2008:CMA} that takes into account 
the instantaneously acting hydrodynamic forces from the fluid on the rigid body. 
A detailed description of the algorithms and methods used in the simulation systems can be found in Pickl \textit{et 
al.}~\cite{Pickl:2012:JOCS, Pickl:2014:IOS}.
%
%
%
\section{Stokes regime: velocity and stroke}
\label{sec:ResultsStokes}
To first test the suitability of the LBM simulation system to reproduce swimming within the Stokes regime, we compare the simulation results with purely Stokesian theory. We have described this theory in previous work~\cite{Pande:2015:SoftM}, where we have determined the velocity of the swimmer 
by considering the effect on each sphere of the different forces it faces, these being the spring forces, the driving forces
and the hydrodynamic forces. For small driving forces, the oscillations of the spheres are small too, and one obtains a coupled system of ordinary differential equations for the sphere positions as functions of time, due to the linear velocity-force relationship in Stokes flow. This system is solved in a perturbative manner with the oscillation of the swimmer arms being the variable of perturbation, and this leads to a swimming velocity given by (in the notation employed in this paper)~\cite{Pande:2015:SoftM}
\begin{equation}\label{eq:v_force}
\mathbf v_\mathrm{force} = \dfrac{7 r \left[2\left(a^2 - b^2\right)\kappa r - a b\left(\kappa^2 + 12r^2\right)\sin\alpha \right]}{24 l^2 \pi^2 \nu^2 \rho^2 \omega \left(\kappa^4 + 40\kappa^2 r^2 + 144 r^4\right)} \hat{\mathbf z}\,.
\end{equation}
Here $\mathbf v_\mathrm{force}$ denotes the swimming velocity calculated using the forces acting upon the swimmer, $\nu$ is the kinematic viscosity of the fluid, $\rho$ is its density, and $\kappa$ is a constant defined for convenience as $\kappa~=~k/(\pi \nu \rho \omega)$. This calculation assumes that there is no slip between the spheres and the fluid, and that the distances between the spheres are much larger than the other length scales in the problem.

In response to the driving forces, the swimming stroke induced is sinusoidal, and can be expressed as
\begin{align}\label{eq:stroke}
\begin{split}
L_1(t) &= l + d_1 \cos(\omega t + \delta_1),\\
L_2(t) &= l + d_2 \cos(\omega t + \delta_2),
\end{split}
\end{align}
where $L_i$ is the instantaneous length of the $i^\text{th}$ arm, and $d_i$ and $\delta_i$ are the  
amplitude and the phase of its oscillation. Once this form of the swimming stroke is adopted, and the earlier-mentioned 
conditions of no fluid slip, large sphere separations and small arm-length oscillations are assumed, then the 
swimmer's velocity can be written as~\cite{Golestanian:2008:PRE}
\begin{equation}\label{eq:v_Gol}
\mathbf v_\mathrm{stroke} = G d_1 d_2 \omega \sin\left(\delta_1 - \delta_2\right) \hat{\mathbf z}.
\end{equation}
Here $\mathbf v_\mathrm{stroke}$ is the swimming velocity (with the subscript referring to the stroke-dependence of its calculation), and $G$ is a geometrical constant which for equal sphere radii $r$ and equal mean arm-lengths $l$ is given by $G = 7r/(24 l^2)$.

The two approaches of finding the swimming velocity--\textit{i.e.}\ assuming known driving forces and known strokes--may be reconciled by determining the various stroke parameters for the assumed force protocol, and then using the now-known strokes to find the velocity as in Eq.~(\ref{eq:v_Gol}). The stroke parameters are found to be
\begin{equation}
\label{eq:stroke_params1}
\begin{split}
d_1 &= \sqrt{\dfrac{a^2 \kappa^2 + 4r^2\left[4a^2 + b^2 - 4 a b \cos\alpha\right] + 4 a b \kappa r \sin\alpha}{\pi^2 \nu^2 \rho^2 \omega^2\left(\kappa^4 + 40 \kappa^2r^2 + 144 r^4\right)}},\\
d_2 &= \sqrt{\dfrac{b^2 \kappa^2 + 4r^2\left[a^2 + 4b^2 - 4 a b \cos\alpha\right] - 4 a b \kappa r \sin\alpha}{\pi^2 \nu^2 \rho^2 \omega^2\left(\kappa^4 + 40 \kappa^2r^2 + 144 r^4\right)}}.\\
\end{split}
\end{equation}
The stroke phase difference $\sin(\delta_1 - \delta_2)$ is presented in the Supplementary Information (SI) due to the heft of the expression involved.

Recasting now the stroke-based velocity form of Eq.~(\ref{eq:v_Gol}) into the force-based one of Eq.~(\ref{eq:v_force}), we find
\begin{equation}
{\mathbf v_\mathrm{stroke}} \thickapprox \dfrac{7(4 l - 3 r)(4 l - 7 r)}{4l(28 l -45r)} 
{\mathbf v_\mathrm{force}}.
\end{equation}
Clearly, to the lowest order in $r/l$, the two velocities match each other perfectly.

\section[Relaxation of inertial swimmers]{Relaxation of inertial swimmers}
\label{sec:inertia_relaxation}
%
%
We wish to observe and identify non-inertial effects in our swimmer's motion by considering its relaxation in the fluid when all forces on it cease. For this purpose, we make the beads in the swimmer much heavier than the surrounding fluid, and let the swimming motion be much faster, with the expectation that both these contributions aid in violating the assumptions of Stokes flow. We therefore run simulations of swimmers with spherical beads using the \Walberla{}-\pe{} framework, where the radius of each sphere is $6 \Delta x$ and the rest length of each arm is $32 \Delta x$. The cycle period is kept at $8000$ time steps to aid simulation accuracy. The phase shift $\alpha$ is set to $0.5\pi$.

The simulations are run in four sets, with the mass of each sphere in the four sets being $20000$, $30000$, $40000$ and $50000$ lattice units, respectively. Note that these masses imply that the sphere densities are about 22 to 55 times larger than the density of the surrounding fluid, respectively, in the four sets, meaning that the beads are not neutrally buoyant. This is not a problem since gravity plays no role in our simulations.

Within each simulation set we increase the driving force amplitudes successively and check first the swimming veolcity of the swimmer in the steady state, and then its relaxation to zero velocity when the driving forces are suddenly switched off. The range of driving force amplitudes checked is slightly different in the different sets, because (as we shall show) the critical force amplitude values where non-inertial effects first become visible depend on the swimmer's mass. The force ranges chosen in each simulation set are those at which the Stokes regime, the non-Stokes regime, and an intermediate regime each become visible. 

\begin{figure}
\centering
  \includegraphics[width=0.7\textwidth]{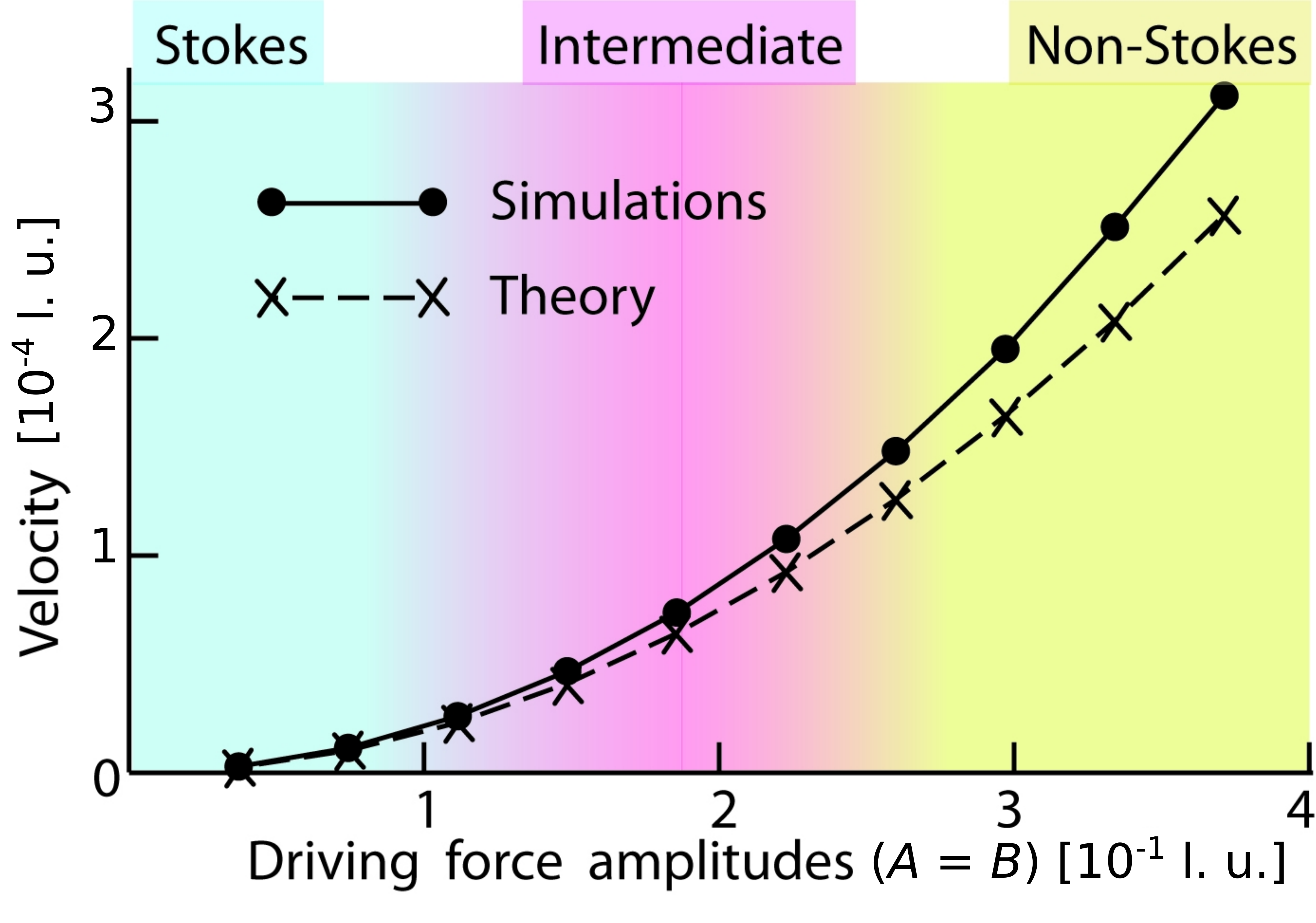}
\caption[Velocity of a swimmer with inertia.]{Velocity of a swimmer with inertia for different force amplitudes, from simulations and theory.}
\label{fig:vel_NonStokes}       
\end{figure}

We find that for all the simulation sets, the Stokesian theory of section \ref{sec:ResultsStokes} gives the correct swimmer velocities in the steady state only for low driving force amplitudes, and diverges from the simulations as the forces increase. In each case, the velocities found from the simulations are higher than those predicted by the theory. As an example, Fig.~\ref{fig:vel_NonStokes} shows the velocity curves from the Stokesian theory and the simulations for a sphere mass of $40000$ lattice units. 
This matches the expectation of a continuous transition from the Stokes regime to the non-Stokes one, which we have loosely marked as blue and yellow regions respectively in Fig.~\ref{fig:vel_NonStokes}. 
In between lies an intermediate regime, marked in pink in Fig.~\ref{fig:vel_NonStokes}, where the transition occurs. We now locate this transition more precisely.
\subsection{Underdamped relaxation of the swimmer}
Our bead-spring swimmer can be viewed as a system of connected harmonic oscillators, which are coupled through both the middle sphere and the surrounding fluid, as well as damped by the fluid. 
Since the drag force is dominant in Stokes flow, 
we postulate 
that the Stokes and the non-Stokes regimes are characterised by overdamping and underdamping, respectively, in the motion of this oscillator system. For an underdamped driven harmonic oscillator, the trajectory is given by
\begin{equation}\label{eq:underdamped}
x(t) = x_0 e^{-\gamma t} \cos\left(\omega_\mathrm{r} t - \alpha_\mathrm{r}\right),
\end{equation}
where $x_0$ is the maximum amplitude, $\gamma$ is the damping constant, and $\omega_\mathrm{r}$ and $\alpha_\mathrm{r}$ are constants of oscillation. For underdamped motion of a coupled system such as ours, and of any mechanical microswimmer in general, it is difficult to specify the different parameters in Eq.~(\ref{eq:underdamped}), yet the damping coefficient $\gamma$ may still be identified. To do this, consider the swimmer's relaxation from an initially steady state when the driving forces are switched off. In the steady state, the swimmer faces the Stokes drag force $F_\mathrm{St} = -6 \pi \eta r_\mathrm{eff} u$ where $r_\mathrm{eff}$ is its effective hydrodynamic radius. When the driving forces vanish, then the body stops instantly if inertial effects are discounted, but in the presence of inertia the body exhibits coasting, and its velocity decreases continuously as
\begin{align}\label{eq:urelax}
F_\mathrm{St} &= m \mathrm du/\mathrm dt = -6 \pi \eta r_\mathrm{eff} u.\nonumber\\
\Rightarrow u &= C e^{-\gamma t}\text{, with } \gamma = \dfrac{6 \pi \eta r_\mathrm{eff}}{m}\text{ and }C\text{ a constant.}
\end{align}
Due to the use of the Stokes drag force $F_\mathrm{St}$ in obtaining $\gamma$, Eq.~(\ref{eq:underdamped}) with said $\gamma$ describes the relaxation only in the intermediate regime between the Stokes and the non-Stokes ones. 
Therefore, by fitting the relaxation curve of the swimmer with Eq.~(\ref{eq:underdamped}), the intermediate regime can be identified. 
This can then also be used to determine the swimmer's effective hydrodynamic radius $r_\mathrm{eff}$, by using Eq.~(\ref{eq:urelax}) to find $r_\mathrm{eff}$ once $\gamma$ has been identified from the fit to Eq.~(\ref{eq:underdamped}). 
This procedure can be used for other mechanical microswimmers such as in \cite{Purcell:1977:AmJPhys, Avron:2005:NewJPhys} which can be viewed as oscillators.
\begin{figure*}
\centering
  \includegraphics[width=0.98\textwidth]{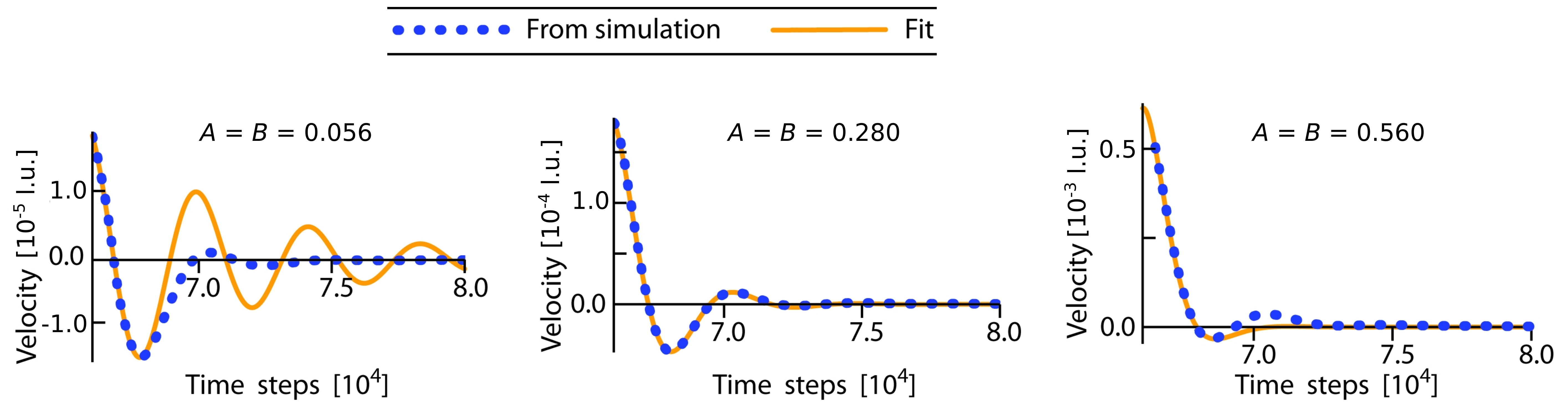}
\caption[Relaxation of a swimmer with inertia.]{Relaxation of a swimmer for different driving force amplitudes.}
\label{fig:relaxation}       
\end{figure*}

\subsection{Identification of intermediate regime}

We now check the relaxation curves obtained from the simulations of our swimmers when the driving forces are turned off after the steady state has been reached. 
We find that Eq.~(\ref{eq:underdamped}) at times underestimates and at times overestimates the damping seen from the simulations. As illustration, Fig. \ref{fig:relaxation} shows the relaxation obtained from simulation (blue dotted curve) and the fit to this curve using Eq.~(\ref{eq:underdamped}) (orange solid curve) for three different force amplitude values for the simulation set with sphere mass = $40000$ lattice units. In each case, the fit parameters are chosen such that they minimise the error to the simulation curve in the initial part of the relaxation, \emph{i.e.}\ from the initial point on the left to the first local minimum. It may be seen that the middle simulation, with a force amplitude of 0.28 lattice units, shows near-perfect agreement with the theoretically predicted relaxation curve based on Eq.~(\ref{eq:underdamped}). In contrast, in the plot on the left, with a force amplitude of 0.056 lattice units, the swimmer's damping is underestimated by the fitting curve, while in the plot on the right with a force amplitude of 0.56 lattice units, the fitting curve from Eq.~(\ref{eq:underdamped}) overestimates the damping. 

The different extents of agreement between the simulations and the theoretical fitting curves for different force amplitudes are understandable, since the theoretical fit is only expected to work well in the regime which shows characteristics of both Stokesian and non-Stokesian motion (due respectively to Eq.~(\ref{eq:urelax}) which depends on the Stokes drag law and Eq.~(\ref{eq:underdamped}) which assumes underdamped motion which is a non-Stokesian effect). This means that the relaxation from the simulation should match the theory the most perfectly in the intermediate regime. Such a reasoning shows the three simulations plotted in Fig.\ref{fig:relaxation} to lie, from left to right, in the Stokes, intermediate, and non-Stokes regimes, respectively.

\begin{figure}
\centering
  \includegraphics[width=0.8\textwidth]{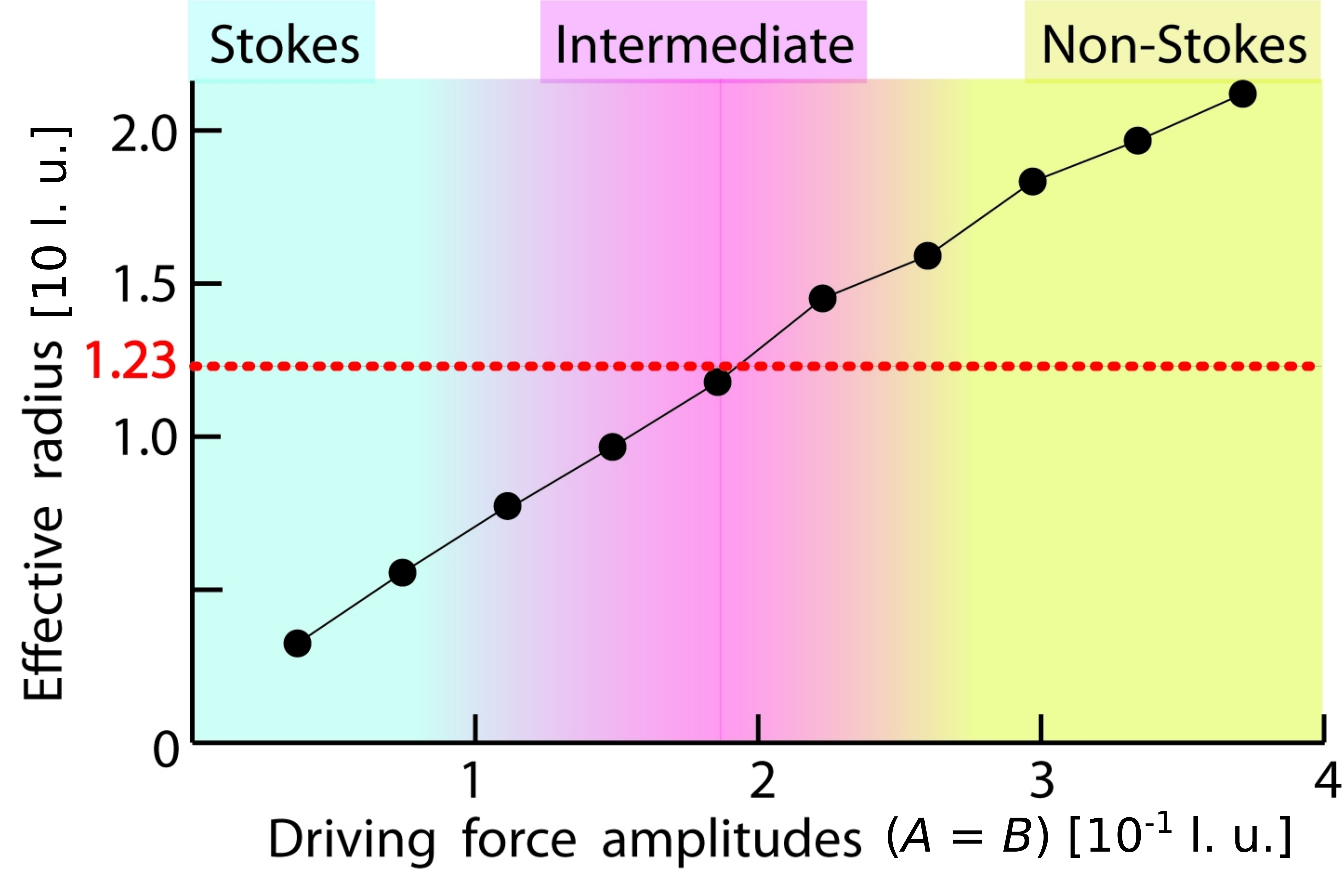}
\caption[Effective radius of a swimmer for different force amplitudes.]{Effective radius of a swimmer for different force amplitudes, as determined by relaxation after force cut-off. Only the simulations in the intermedite regime are expected to reproduce the theoretically predicted value of $12.3$ lattice units (marked by a dashed red line).}
\label{fig:r_eff_m40K}       
\end{figure}

There is another way to check such a determination of the three regimes, by checking the effective radius $r_\text{eff}$ of our swimmer, calculated in Appendix A. In all the simulations that we are here considering, we have $\lambda_i = 6 \Delta x$ and $l_i = 32 \Delta x$, which gives a theoretically expected $r_\text{eff}$ value of $r_\text{eff} = 12.3 \Delta x$. The corresponding values for each simulation can be found by combining Eqs.~(\ref{eq:underdamped}) and (\ref{eq:urelax}). Focussing first on the case of sphere mass = $40000$, we plot the $r_\text{eff}$ value obtained from each simulation in Fig.~\ref{fig:r_eff_m40K}, with the black circles. The theoretical value of $12.3 \Delta x$ is marked with the horizontal dashed red line. It is clear that there is an excellent agreement between the theoretical and the simulation values, for a force amplitude of 0.28 lattice units, \emph{i.e.}\ precisely the force amplitude which we identified as marking the intermediate regime on the basis of the accuracy of the underdamped oscillator description of the swimmer's relaxation. In general, the $r_\mathrm{eff}$ value found from simulations increases monotonically as the driving forces increase, because the parameter $\gamma$ found from Eqs.~(\ref{eq:underdamped}) and (\ref{eq:urelax}) initially underestimates and then overestimates the actual damping in the simulations. This is a good check of our reasoning to identify the intermediate regime.

\begin{figure}
\centering
  \includegraphics[width=0.8\textwidth]{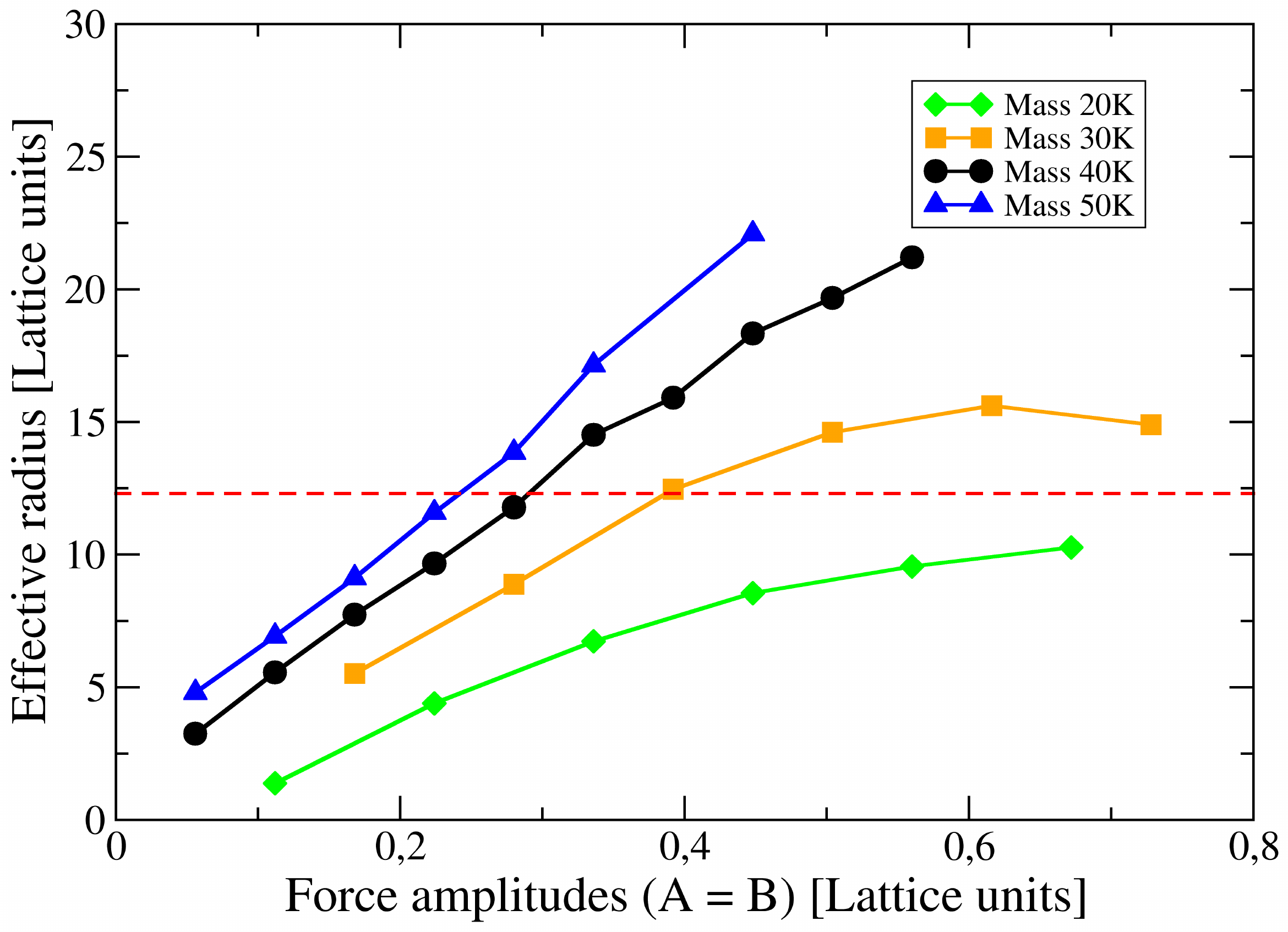}
\caption[Effective radius of a swimmer for different force amplitudes and different mass values.]{Effective radius of a swimmer for different force amplitudes and different mass values, as determined by relaxation after force cut-off.}
\label{fig:r_effs_all}       
\end{figure}

We now consider similar curves obtained from all four sets of simulations. Fig.~\ref{fig:r_effs_all} shows that in each case, the $r_\text{eff}$ values given by the simulations increase monotonically. (The only exceptions to this are the simulations with the very highest force amplitude values probed--see for instance the orange curve for sphere mass = $30000$ lattice units--but that is a relic of the spheres almost colliding with each other at these high driving force values.) Moreover, as the sphere mass increases, the curves shift to the left, and so do the respective force ranges marking the intermediate regime. In each case, the simulations where Eq.~(\ref{eq:underdamped}) best fits the swimmer relaxation are the same ones where the $r_\text{eff}$ curves are closest to the theoretical value, marked by the horizontal red dashed line in Fig.~\ref{fig:r_effs_all}. What this means is that as the mass of the swimmer increases, the non-inertial effects, which first make their presence felt in the intermediate regime, become visible at smaller driving force amplitudes, \emph{i.e.}\ for smaller swimming strokes. This is understandable, since in the Reynolds number it is the product of the velocity and the density which comes together, and increasing either sufficiently breaks the bounds of the Stokes flow assumption. Note that to find the true Reynolds number of a swimmer which is not neutrally buoyant, one should replace the density of the fluid in the definition of the Reynolds number by that of the swimmer itself.

\section[Theory for inertial swimming]{Theory for inertial swimming}
\label{sec:inertia_swimming}

Having phenomenologically studied in the previous section the motion of our swimmer when inertial effects start becoming visible, we now present a basic theoretical treatment of the situation. The calculations are based on our theory for a swimmer with rigid beads swimming in Stokes flow \cite{Pande:2015:SoftM}, and non-Stokesian effects are included in the model by adding a mass acceleration term to the governing equations of motion. As we shall show, the theory provides results which are in good agreement with the simulations of the swimmer with non-negligible mass, and also confirms the three regimes and their characteristics discussed in the previous section.

\subsection{Calculation of velocity for a swimmer with inertia}
\label{sec:inertia_calculation}

We wish to calculate to the lowest non-negligible order the inertia-induced effect on the motion of our bead-spring swimmer. To begin with, we adopt the same model as in section \ref{sec:Model}, with the sinusoidal driving forces specified by Eqs.~(\ref{eq:driving1})-(\ref{eq:driving3}). 
The equation of motion of the $i^\text{th}$ bead can in this case be written as
\begin{equation}\label{eq:v_i_inverse}
\mathbf F^\text{d}_i(t) + \mathbf F^\text{s}_i(t) = \sum\limits_{j = 1}^3 Q_{ij}(t) \mathbf v_j(t),
\end{equation}
where $\mathbf F^\text{d}_i(t)$ and $\mathbf F^\text{s}_i(t)$ are respectively the driving and the spring force on the $i^\text{th}$ bead, and the introduced variables $Q_{ij}(t)$ are functions of the differences $\mathbf R_j(t) - \mathbf R_k(t)\ (j, k = 1, 2, 3)$ of the bead positions. The left hand side in Eq.~(\ref{eq:v_i_inverse}) is the sum of the forces on the $i^\text{th}$ bead not counting the force applied by the fluid, and in the Stokesian description this sum is exactly balanced by the hydrodynamic force of the fluid which opposes the motion of each bead. This is manifestly not the case in the non-Stokes regime, and we postulate that in the latter regime the effect of the applied forces (the driving and the spring forces) on each bead is to accelerate the bead, in addition to neutralising the opposing force provided by the fluid which is given by the right hand side of Eq.~(\ref{eq:v_i_inverse}). In other words, 
\begin{equation}
\mathbf F^\text{d}_i(t) + \mathbf F^\text{s}_i(t) = \sum\limits_{j = 1}^3 Q_{ij}(t) \mathbf v^\text{mass}_j(t) + m_i \dot{\mathbf v}^\text{mass}_i(t),
\end{equation}
where $m_i$ is the mass of the $i^\text{th}$ bead and the velocity of the $i^\text{th}$ bead is now written as $\mathbf v^\text{mass}_i(t)$, in order to highlight its mass-dependence.

Since the sum of the driving and the spring forces over the entire swimmer is still zero, the above equation yields
\begin{equation}\label{eq:force_balance_inertia}
\sum\limits_{i = 1}^3\sum\limits_{j = 1}^3 Q_{ij}(t) \mathbf v^\text{mass}_j(t) + \sum\limits_{i = 1}^3m_i \dot{\mathbf v}^\text{mass}_i(t) = 0.
\end{equation}
Eq.~(\ref{eq:force_balance_inertia}) represents a set of coupled homogenous ordinary differential equations for the velocities $\mathbf v^\text{mass}_i (t)$ of the three beads, which can be solved if a suitable form for the deformations of the lengths of the two arms of the swimmer is given, a specification which allows one to treat the coefficients $Q_{ij}(t)$ as known functions of time. It may be noted that this approach is stroke-based, unlike our force-based approach for the Stokes regime case in section \ref{sec:ResultsStokes}. The form for the armlength deformations that we assume is sinusoidal, given by
\begin{align}\label{eq:stroke_inertia}
\mathbf R_2(t) - \mathbf R_1(t) & = L_1(t) \mathbf{\hat{z}} = \left(l_1 + d_1 \cos(\omega t + \delta_1)\right) \mathbf{\hat{z}},\text{ and}\nonumber\\
\mathbf R_3(t) - \mathbf R_2(t) & = L_2(t) \mathbf{\hat{z}} = \left(l_2 + d_2 \cos(\omega t + \delta_2)\right) \mathbf{\hat{z}}.
\end{align}
Here $\omega$ is the frequency of the swimming cycles (which equals the frequency of the driving forces). The above form for the arm length deformations is the same as that adopted by the arms of the swimmer in the non-inertial (Stokesian) case, where it emerges as a response to the sinusoidal driving forces, and also identical to the form assumed in \cite{Golestanian:2008:PRE} in the stroke-based formulation of the swimmer model in the non-inertial case (Eq.~(\ref{eq:stroke})). In the inertial case the armlengths may be expected to have a dependence on the masses $m_i$, but comparison with the armlength trajectories obtained in simulations suggests that the functions in Eq.~(\ref{eq:stroke_inertia}) describe the armlengths well even in the inertial swimming case if the three beads are identical.

The above equations can be solved fully for unequal masses $m_i$, but here we present the result only for the case $m_i = m$ for the sake of brevity. Using the set of equations (\ref{eq:stroke_inertia}) to specify the coefficients $Q_{ij}(t)$, Eqs.~(\ref{eq:force_balance_inertia}) can be decoupled to
\begin{equation}\label{eq:v_mass_i}
\dot v^\text{mass}_i(t) + \beta(t) v^\text{mass}_i(t) + \gamma_i(t) = 0,
\end{equation}
where $v^\text{mass}_i(t)$ denotes the magnitude of $\mathbf v^\text{mass}_i(t)$, and the coefficients $\beta(t)$ and $\gamma_i(t)$ are given by
\begin{align}
\beta(t) &= \dfrac{1}{3m}\sum\limits_{i = 1}^{3}\sum\limits_{j = 1}^3Q_{ij}(t),\text{ and}\label{eq:beta_inertia}\\
\left[
 \begin{array}{c}
\gamma_1(t)\\
\gamma_2(t)\\
\gamma_3(t)
\end{array}
\right]
 &= \dfrac{1}{3m}
\left[
 \begin{array}{c}
\dot L_{1}(t) \sum\limits_{j = 1}^3 Q_{2j}(t) + \left(\dot L_{1}(t) + \dot L_2(t)\right) \sum\limits_{j = 1}^3 Q_{3j}(t) + m \left(2\ddot L_{1}(t)+\ddot L_{2}(t)\right)\\ 
-\dot L_{1}(t) \sum\limits_{j = 1}^3 Q_{1j}(t) + \dot L_2(t) \sum\limits_{j = 1}^3 Q_{3j}(t) + m \left(-\ddot L_{1}(t) + \ddot L_{2}(t)\right)\\ 
-\left(\dot L_{1}(t) + \dot L_2(t)\right) \sum\limits_{j = 1}^3 Q_{1j}(t) - \dot L_{2}(t) \sum\limits_{j = 1}^3 Q_{2j}(t) - m \left(\ddot L_1(t) + 2\ddot L_2(t)\right)
\end{array}
\right].\label{eq:gamma_i_inertia}
\end{align}

The equations (\ref{eq:v_mass_i})-(\ref{eq:gamma_i_inertia}) are closed by requiring the motion to be periodic in the steady state, \emph{i.e.}
\begin{equation}\label{eq:v_periodic}
v^\text{mass}_i(t) = v^\text{mass}_i(t + 2\pi/\omega).
\end{equation}
In terms of the velocities of the individual beads, the velocity $\mathbf v^\text{mass}$ of the whole swimmer is
\begin{equation}\label{eq:v_swimmer_inertia}
\mathbf v^\text{mass} = v^\text{mass} \mathbf{\hat{z}} = \frac{\omega}{6\pi}\int\limits_{0}^{2\pi/\omega}\sum\limits_{i=1}^{3}\mathbf v^\text{mass}_i(t) \mathrm dt,
\end{equation}
where the averaging of the bead velocities has been done over the three beads and over one swimming cycle.

Eqs.~(\ref{eq:v_mass_i})-(\ref{eq:v_swimmer_inertia}) are solved for $\mathbf v^\text{mass}$ by the integrating factor method, and the final closed-form expression comes out to be
\begin{align}
v^\text{mass} = &\dfrac{\omega}{6\pi}\int\limits_{0}^{2\pi/\omega}\sum\limits_{i=1}^{3}\dfrac{1}{I(t)} \left[-\int\limits_{0}^{t}\gamma_{i}(t') I(t') \mathrm dt' + \dfrac{\dfrac{1}{I(2\pi/\omega)}-\int\limits_{0}^{2\pi/\omega}\gamma_{i}(t')I(t') \mathrm dt'}{1-\dfrac{1}{I(2\pi/\omega)}}\right]\mathrm dt, 
\end{align}
where the function
\begin{equation}
I(t) = \exp\left(\int\limits_{0}^{t}\beta(s)\mathrm ds\right)
\end{equation}
has been introduced for ease of expression.

\subsection{Comparison with simulations}
\label{sec:simulation_comparison_inertia}

\begin{figure}
\centering
  \includegraphics[width=0.66\textwidth]{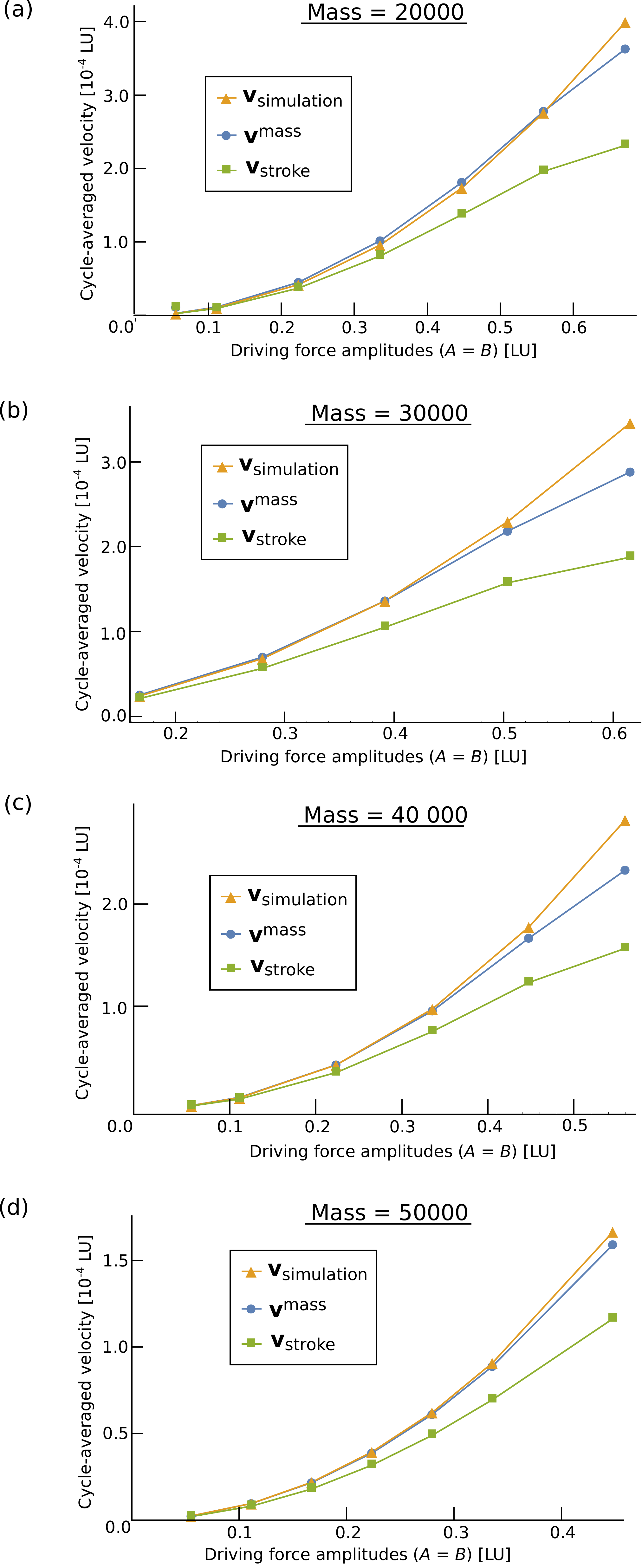}
\caption[Comparison of $\mathbf v_\text{simulation}$, $\mathbf v^\text{mass}$ and $\mathbf v_\text{stroke}$ for different mass values.]{Comparison of velocity expressions from simulations ($\mathbf v_\text{simulation}$) and from theories which do ($\mathbf v^\text{mass}$) and do not ($\mathbf v_\text{stroke}$) include the effect of inertia, for different mass values.}
\label{fig:comparison_v_inertia}       
\end{figure}

We check the results of our calculation for the same four sets of simulations (for sphere masses $20000$, $30000$, $40000$ and $50000$ on the lattice) discussed in section \ref{sec:inertia_relaxation}. Fig.~\ref{fig:comparison_v_inertia} displays the velocity of the swimmer for each mass value as a function of the driving forces. The different velocity curves shown are the one obtained from the simulations (labelled as $\mathbf v_\text{simulation}$ and marked with orange triangles), the one given by $\mathbf v^\text{mass}$ using the theory of section~\ref{sec:inertia_calculation} (marked with blue circles), and the one given by $\mathbf v_\text{stroke}$ in Eq.~(\ref{eq:v_Gol}) which does not consider the effect of inertia (marked with green squares). 

An immediate observation is that the $\mathbf v^\text{mass}$ curves match much better with the $\mathbf v_\text{simulation}$ ones than the $\mathbf v_\text{stroke}$ curves do. The higher the mass the better is the agreement between the $\mathbf v_\text{simulation}$ curves and the theoretical $\mathbf v^\text{mass}$ curves. This shows that at least in the parameter ranges that we have explored, our velocity calculation for swimming with inertia works well.

\begin{figure}
\centering
  \includegraphics[width=0.8\textwidth]{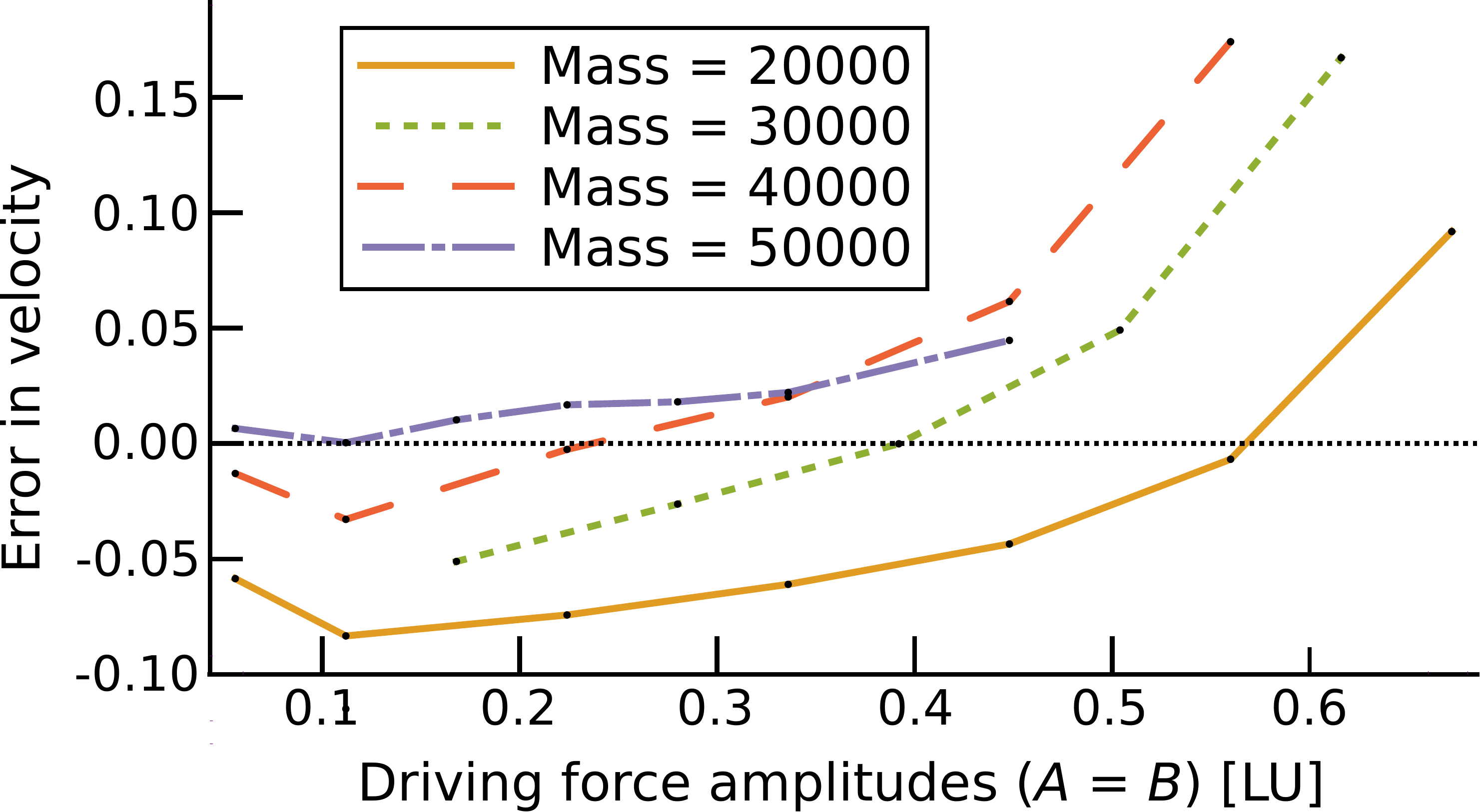}
\caption[Relative error between $\mathbf v_\text{simulation}$ and $\mathbf v^\text{mass}$ for different mass values.]{Relative error between $\mathbf v_\text{simulation}$ and $\mathbf v^\text{mass}$ for different mass values.}
\label{fig:v_inertia_errors}       
\end{figure}

A second result is that the $\mathbf v^\text{mass}$ curves initially overestimate the simulation velocities and then, beyond some force amplitude value dependent on the mass value, begin to underestimate them. This is most clearly seen in parts (a) and (b) of Fig.~\ref{fig:comparison_v_inertia}, for masses $20000$ and $30000$ on the lattice, but is also the case for the other two mass values. To show this more clearly, we plot in Fig.~\ref{fig:v_inertia_errors} the relative errors between the $\mathbf v_\text{simulation}$ and the $\mathbf v^\text{mass}$ curves for the different sphere masses as functions of the driving force amplitudes. Here one sees clearly that the relative errors, never very large ($\lesssim 15\%$), are initially negative and then become positive for each mass value except for mass 50000, in which case they are small enough ($< 5\%$) to be within the error bounds of the LBM. Moreover, the errors get progressively smaller as the mass of the spheres increases.

To sum up, the observations above and in section~\ref{sec:inertia_relaxation} lead to the following conclusions:
\begin{enumerate}
\item The Stokesian approaches for predicting the swimmer's motion work well for low driving forces (or equivalently, for small swimming strokes).
\item As the driving forces/strokes get large, the swimming begins to diverge from the Stokesian predictions.
\item Combining the Stokesian approach with non-Stokesian elements--such as an underdamped relaxation of the swimmer or a mass acceleration term in the equations of motion--leads to accurate description of the swimmer's motion in a narrow, intermediate regime between the Stokes and non-Stokes regimes. To the left of this regime (in a velocity vs.\ driving force plot), the combined approach overestimates the swimmer velocity, and to the right of the regime it underestimates the velocity, which is consistent with the identification of the left and right regions as the Stokesian and non-Stokesian regimes, respectively.
\item As the mass of the swimmer increases, the intermediate regime becomes visible at smaller driving forces (or swimming strokes).
\end{enumerate}

It may be observed that there is a small difference between the values of the driving force amplitudes in Figs.\ref{fig:r_effs_all} and \ref{fig:v_inertia_errors} at which, respectively, the $r_\text{eff}$ curves equal the theoretical value and the errors between the $\mathbf v_\text{simulation}$ and $\mathbf v^\text{mass}$ curves become zero. This is to be expected since the methods of sections $\ref{sec:inertia_relaxation}$ and $\ref{sec:inertia_swimming}$ are quite different, with the one based on the force-free relaxation of the swimmer, and the other on the force-induced swimming. Nevertheless, the fact that both the methods lead to the above listed conclusions points to the validity of the general approach in trying to identify inertial features in the motion.

\section{Appendix A}
For negligibly small oscillation amplitudes, \textit{i.e.}\ to the zeroth order in $d_i/l_j$ (from Eq.~(\ref{eq:stroke})), the $r_\mathrm{eff}$ for the swimmer over a cycle is the same as that of the equilibrium configuration moving rigidly with a constant velocity. 
To obtain such a configuration, we let the three beads have effective friction coefficients $\lambda_1$, $\lambda_2$ and $\lambda_3$, and let constant forces $\lambda_1\mathbf F$, $\lambda_2\mathbf F$ and $\lambda_3\mathbf F$ respectively act on the three beads, with $\mathbf F = F \mathbf{\hat{z}}$. We will first show that in the steady state these forces indeed result in a rigid equilibrium configuration, \emph{i.e.}\ the springs are unextended.

Assume that in the steady state both the springs are extended, with the respective extensions being $\Delta l_1$ and $\Delta l_2$. Then, the forces on the left, middle and right spheres, apart from the hydrodynamic forces, are respectively $(\lambda_1 F + k\Delta l_1)$, $(\lambda_2 F + k\Delta l_2 - k\Delta l_1)$, and $(\lambda_3 F - k\Delta l_2)$, all in the +$z$-direction. Since we are considering the steady state, all the three spheres are assumed to be moving with the same speed, which equals $\dfrac{F_i^\text{d+s}}{6\pi\eta \lambda_i}$. Therefore, we have
\begin{equation}
\dfrac{(\lambda_1 F + k\Delta l_1)}{6 \pi \eta \lambda_1} = \dfrac{(\lambda_2 F + k\Delta l_2 - k\Delta l_1)}{6 \pi \eta \lambda_2} = \dfrac{(\lambda_3 F - k\Delta l_2)}{6 \pi \eta \lambda_3}.
\end{equation}
\begin{align}
\Rightarrow & F + \dfrac{k\Delta l_1}{\lambda_1} = F + \dfrac{k \Delta l_2}{\lambda_2} - \dfrac{k \Delta l_1}{\lambda_2} = F - \dfrac{k \Delta l_2}{\lambda_3}.\\
\Rightarrow & \dfrac{\Delta l_1}{\lambda_1} = \dfrac{\Delta l_2 - \Delta l_1}{\lambda_2} = -\dfrac{\Delta l_2}{\lambda_3}.
\end{align}
If $\Delta l_1$ is positive (negative), then $\Delta l_2$ must be negative (positive), and then the middle term is negative (positive), resulting in a contradiction. Therefore, the only solution to the above equation is
\begin{equation}
\Delta l_1 = \Delta l_2 = 0
\end{equation}
which is what we wanted to show.

Then the velocity of the swimmer in the steady state is
\begin{align}
\mathbf v_\text{force} & =  \dfrac{1}{3T}\int\limits_0^T\sum\limits_{i = 1}^3 \mathbf v_i \, \mathrm dt\nonumber\\
& = \dfrac{1}{3T}\int\limits_0^T\sum\limits_{i = 1}^3 \left[\dfrac{1}{6\pi\eta \lambda_i}\,\lambda_i\mathbf F + \sum\limits_{j \neq i}^3 \mathbf T\left(\mathbf R_i - \mathbf R_j\right)\cdot\lambda_j \mathbf F\right] \, \mathrm dt.
\end{align}
As the springs are always at their rest lengths, all the terms in the above expression are time-independent, and therefore can be taken out of the time integral. So
\begin{align}
\mathbf v_\text{force} = &\dfrac{\mathbf F}{6\pi\eta} + \dfrac{1}{3}\left[\sum\limits_{i = 1}^3\sum\limits_{j \neq i}^3 \lambda_j \mathbf T\left (\mathbf R_i - \mathbf R_j\right)\right]\cdot\mathbf F\nonumber\\
= &\dfrac{\mathbf F}{6\pi\eta} + \dfrac{1}{3}\left(\lambda_1\mathbf T_{12} + \lambda_1 \mathbf T_{13} + \lambda_2 \mathbf T_{21} + \lambda_2 \mathbf T_{23} + \lambda_3 \mathbf T_{31} + \lambda_3 \mathbf T_{32}\right)\cdot \mathbf F,
\end{align}
where $\mathbf T_{ij}$ is shorthand for $\mathbf T\left(\mathbf R_i - \mathbf R_j\right)$.

Now,
\begin{align*}
& \lambda_1\left(\mathbf T_{12} + \mathbf T_{13}\right)\cdot\mathbf F = \dfrac{1}{8\pi\eta}\left(\dfrac{1}{l_1} + \dfrac{1}{l_1 + l_2}\right)(2\lambda_1\mathbf F).\\
& \lambda_2\left(\mathbf T_{21} + \mathbf T_{23}\right)\cdot\mathbf F = \dfrac{1}{8\pi\eta}\left(\dfrac{1}{l_2} + \dfrac{1}{l_1}\right)(2\lambda_2\mathbf F).\\
& \lambda_3\left(\mathbf T_{31} + \mathbf T_{32}\right)\cdot\mathbf F = \dfrac{1}{8\pi\eta}\left(\dfrac{1}{l_1 + l_2} + \dfrac{1}{l_2}\right)(2\lambda_3\mathbf F).
\end{align*}

Therefore,
\begin{align}
\mathbf v = & \dfrac{\mathbf F}{6\pi\eta} + \dfrac{2\mathbf F}{3}\left(\dfrac{1}{8\pi\eta}\right)\left[\lambda_1\left(\dfrac{1}{l_1} + \dfrac{1}{l_1 + l_2}\right) + \lambda_2\left(\dfrac{1}{l_1} + \dfrac{1}{l_2}\right)\right.\nonumber\\
	& \ \ \ \ \ \ \ \ \ \ \ \ \ \ \ \ \ \ \ \ \ \ \ \ \ \ \ \ \ \ \ \ \ \ \ \left.+ \lambda_3\left(\dfrac{1}{l_1 + l_2} + \dfrac{1}{l_2}\right)\right]\nonumber\\
= & \dfrac{\mathbf F}{6\pi\eta}\left\{1 + \dfrac{1}{2}\left[\lambda_1\left(\dfrac{1}{l_1} + \dfrac{1}{l_1 + l_2}\right) + \lambda_2\left(\dfrac{1}{l_1} + \dfrac{1}{l_2}\right)\right.\right. \nonumber\\
 & \ \ \ \ \ \ \ \ \ \ \ \ \ \ \ \ \ \ \ \ \ \ \ \left . \left . + \lambda_3 \left (\dfrac{1}{l_1 + l_2} + \dfrac{1}{l_2} \right ) \right ] \right \} .
\end{align}

By the definition of $r_\text{eff}$, we have
\begin{align}
\mathbf v =  \dfrac{\mathbf F_\text{total}}{6\pi\eta r_\text{eff}} =  \dfrac{(\lambda_1 + \lambda_2 + \lambda_3)\mathbf F}{6\pi\eta r_\text{eff}},
\end{align}
from where we get
\begin{align}\label{eq:r_eff}
\dfrac{1}{r_\text{eff}} = \dfrac{1}{\lambda_1 + \lambda_2 + \lambda_3}&\left\{1 + \dfrac{1}{2}\left[\lambda_1\left(\dfrac{1}{l_1} + \dfrac{1}{l_1 + l_2}\right) + \lambda_2\left(\dfrac{1}{l_1} + \dfrac{1}{l_2}\right)\right.\right. \nonumber\\
& \ \ \ \ \ \ \ \ \ \ \ \ \ \ \ \ + \left.\left.\lambda_3\left(\dfrac{1}{l_1 + l_2} + \dfrac{1}{l_2}\right)\right]\right\}.
\end{align}

For identical and equidistant beads (for which we have $\lambda_i = \lambda$ and $l_i = l$), the above expression for the swimmer effective radius simplifies to
\begin{equation}\label{eq:r_eff_identical_beads}
r_\text{eff} = \dfrac{6 \lambda l}{2l + 5\lambda}.
\end{equation}
For identical beads, the expression in Eq.~(\ref{eq:r_eff_identical_beads}), correct to the first order in $\lambda/l$ and the zeroth order in $d_i/l$, can also be obtained from the swimmer effective radius provided in Eq.\ (40) in \cite{Golestanian:2008:PRE}.

\bibliography{Literaturliste} 
\bibliographystyle{rsc3}
 \end{document}